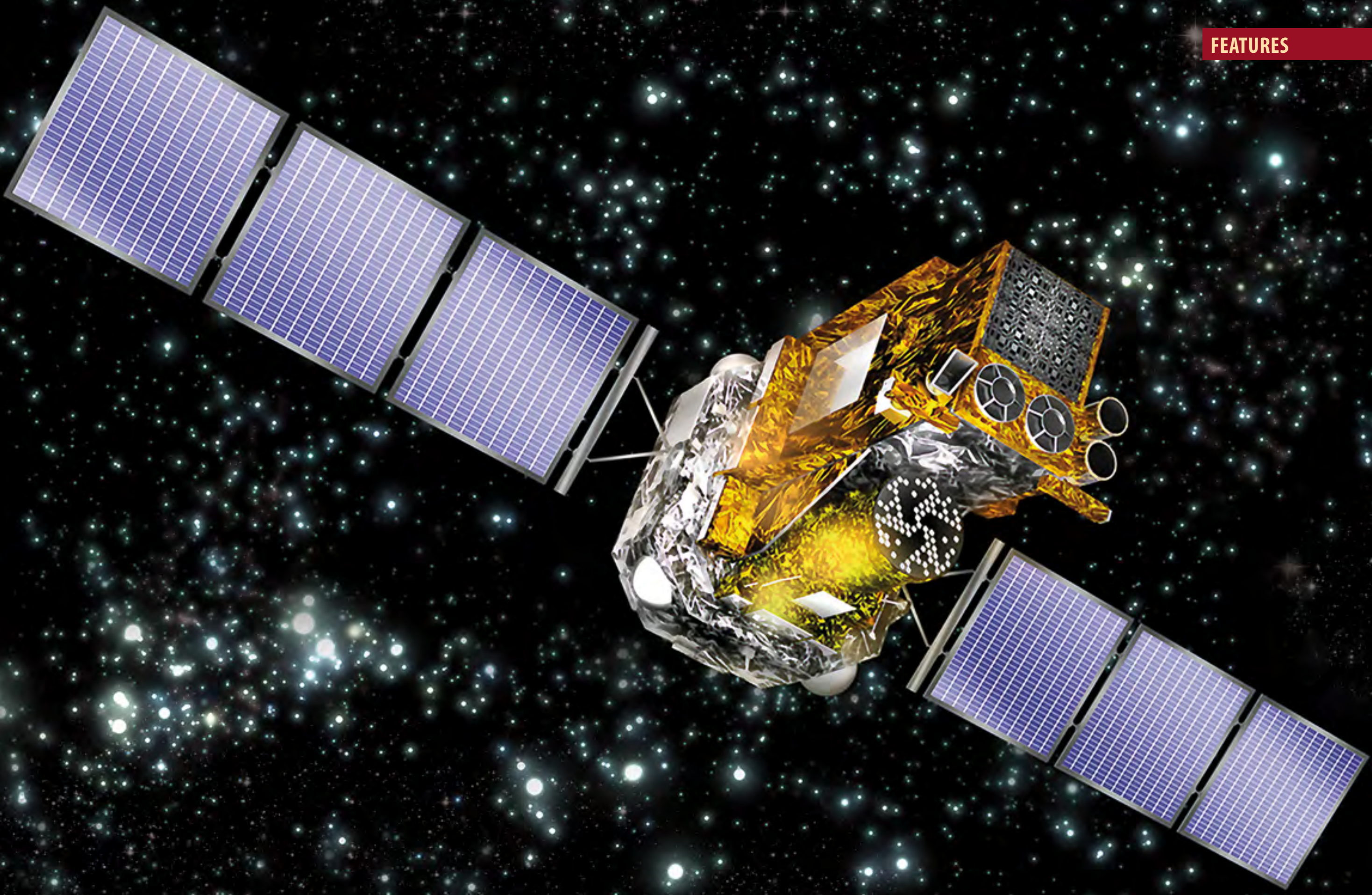

# SURPRISES IN THE HARD X-RAY SKY

Thierry J.-L. Courvoisier – Data Centre for Astrophysics, University of Geneva 

Cosmic objects emit throughout the electro-magnetic spectrum, from radio waves to very high-energy gamma rays. Some spectral regions can be observed from the ground, but space instrumentation is needed for most of them. Some spectral regions are particularly difficult, for example hard X-rays and MeV gamma rays. For one thing, in this spectral region photons interact least with matter. Moreover, no focussing optics was available until very recently. INTEGRAL, the gamma ray observatory mission led by ESA, was designed to observe the sky in this very domain.

Astrophysics space missions are long-term endeavours. INTEGRAL, the gamma ray satellite launched by the European Space Agency (ESA) in 2002 is no exception. The mission was proposed by a group of European and US scientists to ESA in 1989. Following a number of studies and successful competitions it was selected in 1993. The years between 1995 and 2002 were a long succession of strenuous efforts. The first hurdle was the replacement of US and UK funds, which, although formally committed at the time of selection, did not materialise. Then came a series of technical difficulties, most rather unexpected, like major difficulties to glue radiation-hardened components in such a way that they could endure temperature excursions of some 200K. Not trivial either were the discussions with the Soviet-Union, which transformed into Russia during the process, to obtain a launch on a Proton rocket free of financial charge, but for a reasonable share of the data.

All these hurdles were overcome, not least thanks to the competence of all the teams involved and the strong will of all the actors: scientists, engineers and staff from ESA and national funding bodies. The success is demonstrated by the fact that INTEGRAL was originally funded for two years of operations, designed for five years in orbit, was launched on October 17 2002 and is still functioning flawlessly in Summer 2013.

The mission [6] was designed to provide observations with high-energy resolution through a spectrometer, SPI [5], and high angular resolution through an imager, IBIS [4]. "High" is to be understood here in terms of hard X-ray and gamma ray astronomy, and means that the spectral resolution $\lambda/\Delta\lambda$ is of the order of 400 at 1MeV and the angular resolution about 12'. These two core instruments are complemented with an X-ray monitor, JEM-X [2], and an optical camera, OMC [3]. Together these instruments provide a hard X-ray and gamma ray coverage from a few keV to several MeV, a very wide spectral band. The data are collected at the ISDC [1], where they are processed, archived and distributed to the world astronomical community.

## Five outstanding results

Ten years into the mission, a number of results have been collected. As often when the sensitivity of a set of instruments in a spectral domain increases by a large factor, surprising results are obtained. INTEGRAL is no exception to this in the hard X-rays and soft gamma rays. The first surprise was to discover sources that are bright above 10 keV and were unknown as X-ray sources before the INTEGRAL measurement. This is insofar surprising because the INTEGRAL instruments use a coded mask optics (the shadow of a mask is measured and the "shadowgramme" is deconvolved to obtain a sky image) that is considerably less efficient than the focusing optics that can be used below 10 keV. The expectation was thus that INTEGRAL sources would have been known from lower energy X-ray sky surveys. A large collection of results is presented in [7], where extensive bibliographical references can be found. We highlight here five domains in which INTEGRAL observations have made important contributions. The choice reflects the prejudice of the author; others would most probably have highlighted a somewhat different set of domains.

## Populations of galactic compact objects

X-rays are absorbed by dust and gas in interstellar space. The absorption cross-section increases towards lower energies, roughly with the third power of inverse photon energy. Sources embedded in relatively dense environments therefore shine only above 10 keV or so. Since surveys using X-ray focusing telescopes had been sensitive only at energies less than 10 keV, such sources could not be discovered prior to INTEGRAL observations.

INTEGRAL/IBIS has a very large field of view of about 30 degrees and a much improved sensitivity compared with its predecessor instruments. It discovered serendipitously hundreds of hitherto unknown highly absorbed bright X-ray binary sources in our Galaxy (fig. 1). A large fraction of these sources are binary systems in which a neutron star orbits close to a giant star in regions that are dense from the wind emitted by the giant. This dense matter is in part accreted by the neutron star, causing it to radiate the gravitational binding energy of the accreted material as X-rays, and also absorbs the X-rays at the lower energies, letting only the higher energy X-rays shine outwards. The compact neutron star thus serves in a certain way as a probe of the stellar wind conditions in its surrounding, giving a way to study not only the accretion process, but also the wind conditions.

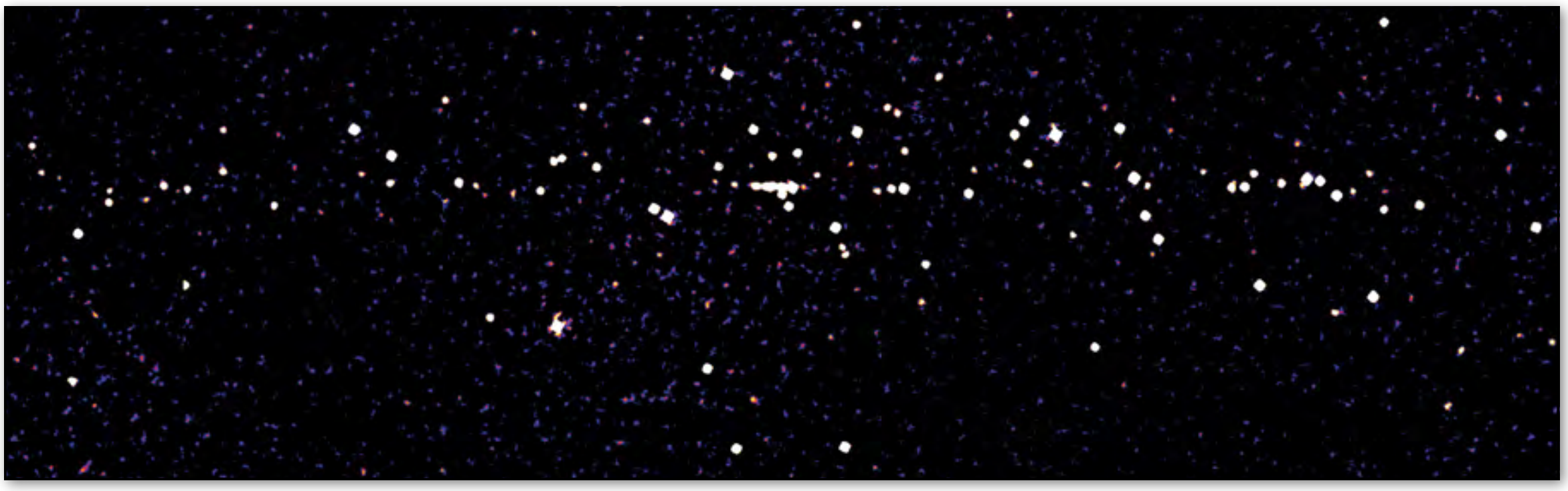

▶ FIG. 1: The central regions of the Galaxy as seen by INTEGRAL (Credit ISDC/ R. Walter).

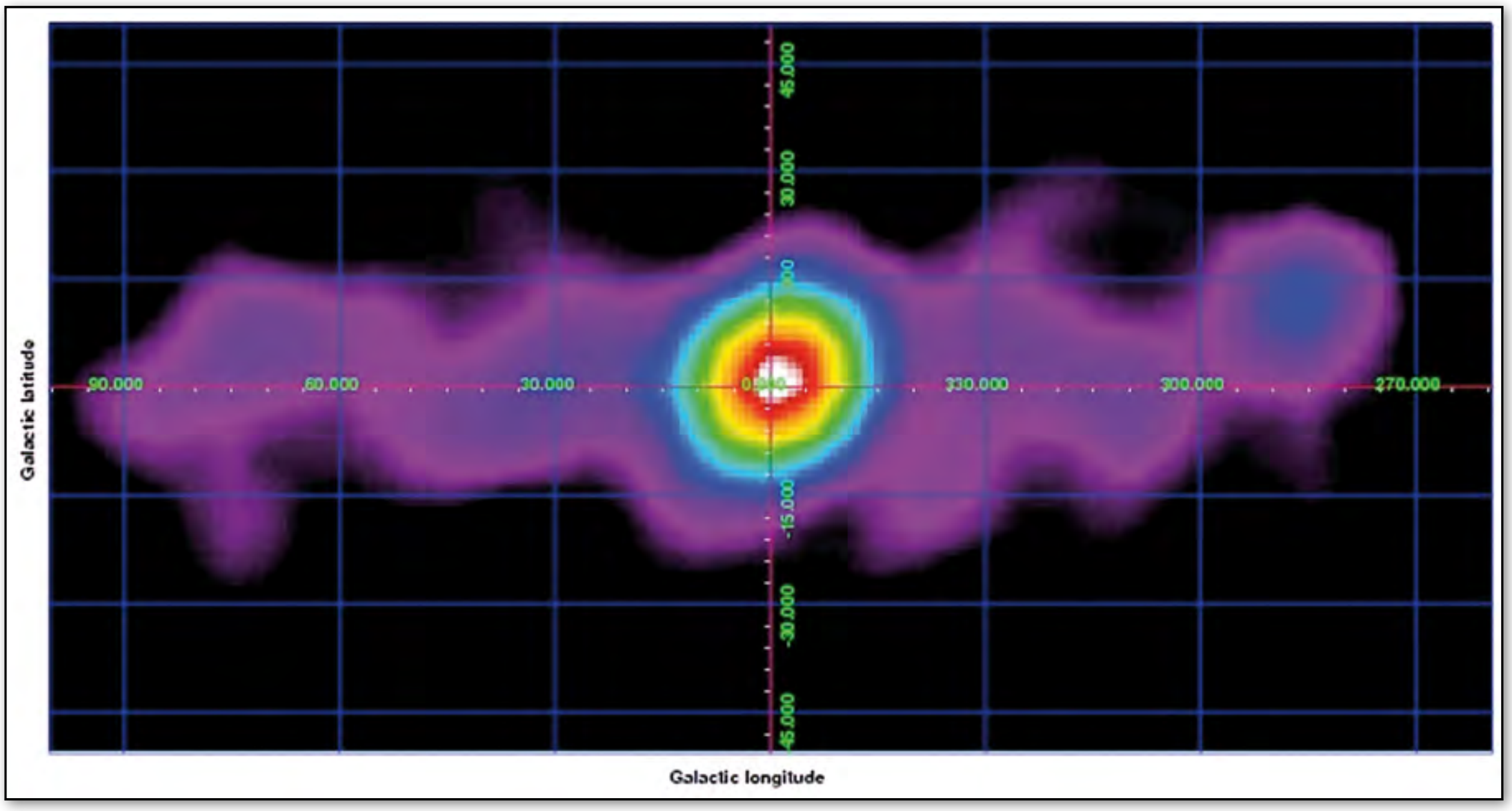

◂ **FIG. 2:** Electron positron annihilation in the central region of the Galaxy (Credit ESA/Bouchet).

## Electron–positron annihilation in the central regions of the Galaxy

It has been known for a number of years that "cold" electrons and positrons annihilate in the central region of the Galaxy. This annihilation gives rise to a narrow emission line at 511 keV, the rest energy of the electron. INTEGRAL/SPI has imaged the line emission in the central region of the galaxy and resolved the line in energy. The temperature of the electrons and positrons can be deduced from the line profile; it is some 5000 K. The SPI image shows that the emission has two components: one spherical centred around the centre of the Galaxy and the other associated with the disk of the Galaxy (fig. 2). While the origin of the electrons poses no difficulty, that of the positrons still eludes our understanding. The disk emission can be due to positrons originating in radio-active decays of elements synthetised in stars. However, no convincing origin has been found for the positrons giving rise to the spherical component. Many explanations have been brought forward in the literature. They range from positrons created in binary systems in the bulge of the Galaxy, to positrons that would be created by radioactive processes in the disk of the Galaxy and transported, possibly along magnetic field lines, to its central regions, to positrons associated with the decay of as yet unidentified dark matter particles. None of these possibilities is free of significant difficulties. The discussion is ongoing.

## Variability in the X-ray sky

One of the most important findings of X-ray astronomy over the last decades is the extreme variability that many sources exhibit. While observations of the sky in visible light has led us to think that stability is the rule, except within the orbit of the Moon, high energy astrophysics has taught us otherwise: violent physical phenomena take place on timescales as short as milliseconds (gamma ray bursts), seconds to hours and days (binary stars and even quasars). The wide field of view of INTEGRAL has allowed observers to catch many unexpected phenomena and to discover a new class of objects, the Supergiant Fast X-ray Transients (SFXT). The outbursts of these sources may last for as little as some hours and take place very rarely, hence the difficulty to catch them in the act. These sources, like the absorbed sources discussed above, are binary systems in which a compact object orbits close to a massive star.

Other types of variable sources detected by INTEGRAL include pulsars in binary systems that accrete matter and angular momentum. The latter accretion could be measured through precise timing observations by an X-ray instrument of a new INTEGRAL source. The phase of the pulsed emission shifted during the few weeks in which the source was active, thus convincingly showing the neutron star spin acceleration.

Very recently the nucleus of a Galaxy brightened considerably for a fraction of a year. This isolated event is due to

▾ **FIG. 3:** Giant flare from the soft gamma repeater 1806-20. The source of the flare was not in the field of view of the INTEGRAL instrument. The flare is, however, so bright that it is recorded in the detector's active shield (credit MPE).

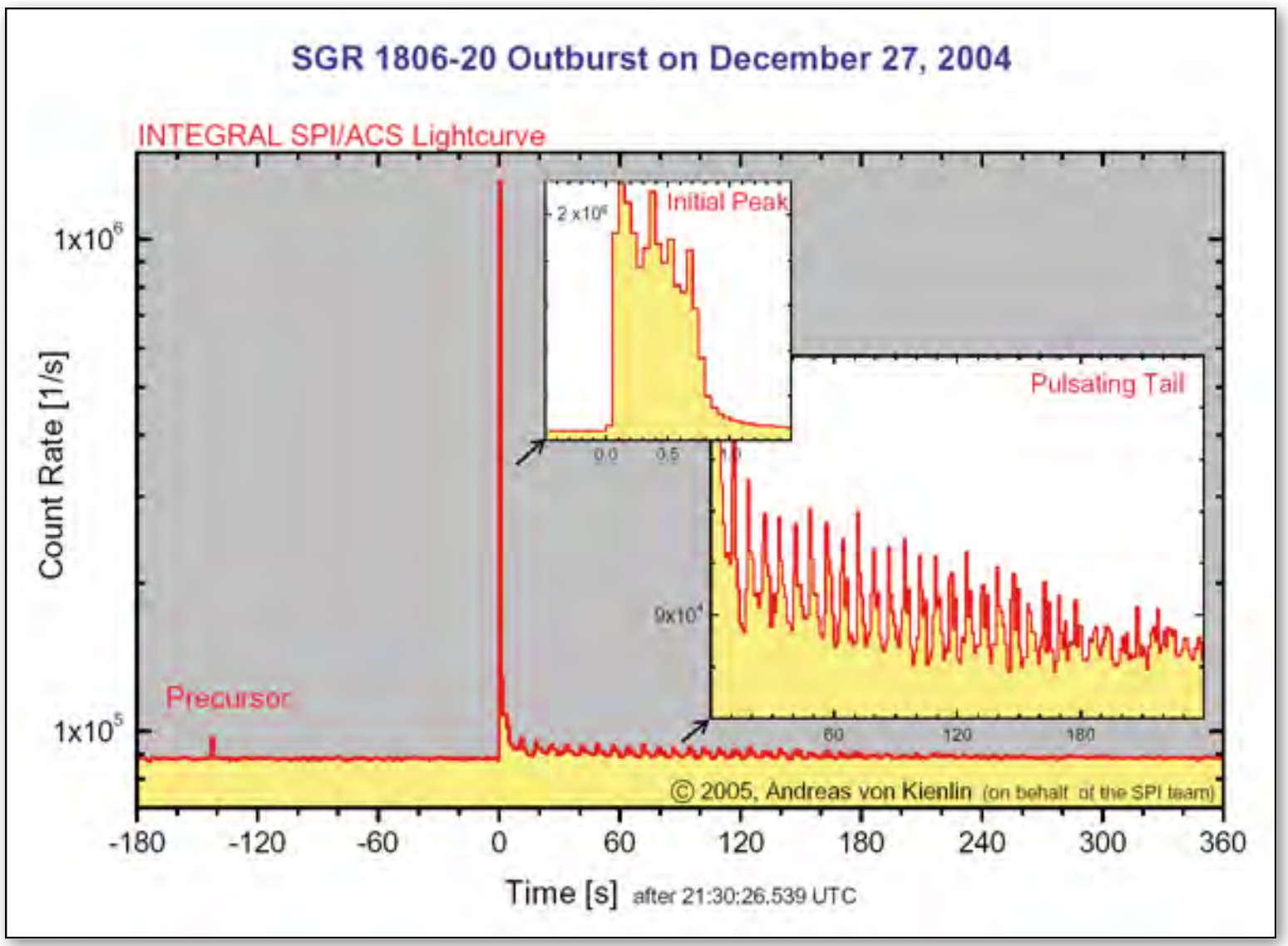

the disruption of a lone massive planet that cruised too close to the massive black hole hosted at the centre of the Galaxy. The planet was tidally disrupted and subsequently accreted onto the black hole releasing gravitational energy as X-ray radiation in the process.

## Magnetars

The energy source in X-ray sources is most often – but not always – due to the matter falling in the deep potential well of a compact object. Some sources in our Galaxy are isolated so that the nature of the energy radiated cannot be due to accretion from a companion. These objects, magnetars, are powered by their magnetic field. They come in two sorts, the anomalous X-ray pulsars and the soft gamma repeaters. Magnetic fields of up to $10^{15}$ Gauss (or $10^{11}$ T) are found in these objects, flares are caused by the reorganisation of field lines as the solitary objects slow down in their rotation (fig. 3). A surprise is that the energy output from these objects is maximum in the hard X-rays, and that this is not a result of local absorption, nor of interstellar absorption but really seems to be caused by the as yet poorly understood emission mechanisms.

## Active galaxies and the X-ray background

In 1962, the first rocket that measured X-rays from space beyond the Sun discovered a bright source, Sco X-1, and a ubiquitous radiation. This was later called the diffuse extra-galactic X-ray background. At low X-ray energies, this background could be resolved in individual weak sources. The origin of this background at higher energies is more difficult to assess. On one side the spectral shape of the background does not match that expected from the superposition of weak extra-galactic sources in our surroundings, and on the other side hard X-ray instruments could not image weak sources. INTEGRAL has been able to resolve some 2.5% of the diffuse X-ray extra-galactic background, and has also shown that the hard X-ray spectral shape of local active galaxies differs from previous expectations. This leads one to think that, with the advent of focusing optics between 10 and 100 keV, this component will indeed also be resolved in individual sources.

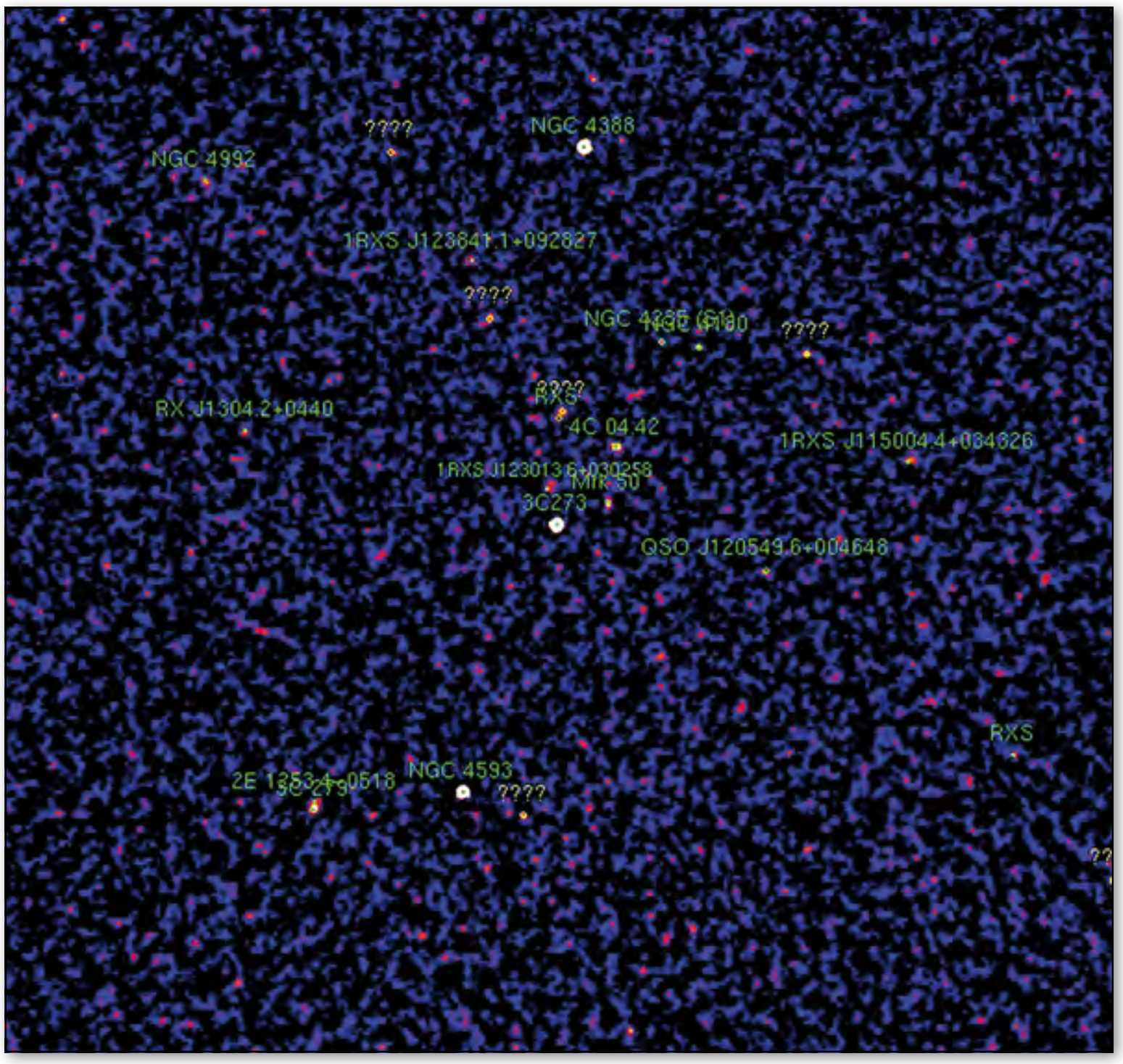

▾ FIG. 4: The deepest INTEGRAL extra-galactic image. Individual sources represent some 2.5% of the diffuse X-ray background.

## Conclusion

INTEGRAL had been funded for two years of observations and designed for five years of orbital life. It is now more than ten years that INTEGRAL functions flawlessly and provides a continuous string of new and often unexpected results. We can only hope that reasonable funding authorities will continue the support to the mission as long as the instruments and satellite function as well as they do now, and new results keep hitting the press.

## About the Author

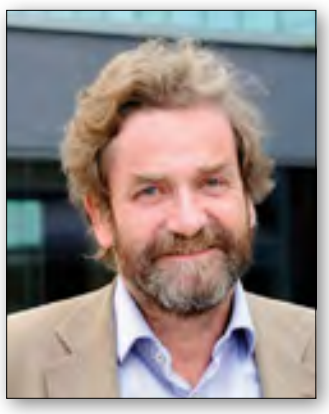

**Thierry J.-L. Courvoisier** studied theoretical physics in Zurich where he obtained a PhD in 1980. He worked at the European Space Operations Centre (ESOC) in Darmstadt and at the European Southern Observatory (ESO) in Garching bei Muenchen before joining the university of Geneva in 1988 where he is full professor. He is the author of "High energy astrophysics, an introduction", Springer 2012.
Thierry J.-L. Courvoisier is president of the Swiss Academy of natural Sciences and of the Swiss Academies of Arts and Sciences. He is also president of the European Astronomical Society (EAS).

## Acknowledgements

It has been a rare privilege to be associated with the INTEGRAL mission for some 25 years. The results presented here stem from the astronomical community at large that makes an extensive use of the satellite. Their work rests on the efforts of the teams behind the instruments and the ISDC. My deep gratitude goes to all who shared their knowledge and results with me in the course of the years.